\documentclass[aps,prd,preprint,superscriptaddress]{revtex4-1}

\usepackage{graphicx}

\begin{document}


\title{A theory of static friction between homogeneous surfaces based on compressible elastic smooth microscopic inclines}


\author{Freeman Chee Siong Thun}
\email{thun0002@e.ntu.edu.sg}
\affiliation{Division of Physics and Applied Physics, School of Physical and
Mathematical Sciences, Nanyang Technological University, 21
Nanyang Link, Singapore 637371}

\author{Vee-Liem Saw}
\email{VeeLiem@ntu.edu.sg}
\affiliation{Division of Physics and Applied Physics, School of Physical and
Mathematical Sciences, Nanyang Technological University, 21
Nanyang Link, Singapore 637371}

\author{Kin Sung Chan}
\email{ch0085ng@e.ntu.edu.sg}
\affiliation{Division of Physics and Applied Physics, School of Physical and
Mathematical Sciences, Nanyang Technological University, 21
Nanyang Link, Singapore 637371}


\date{\today}

\begin{abstract}
We develop a theory of static friction by modeling the homogeneous surfaces of contact as being composed of a regular array of compressible elastic smooth microscopic inclines. Static friction is thought of as the resistance due to having to push the load over these smooth microscopic inclines that share a common inclination angle. As the normal force between the surfaces increases, the microscopic inclines would be compressed elastically. Consequently, the coefficient of static friction does not remain constant but becomes smaller for a larger normal force, since the load would then only need to be pushed over smaller angles. However, a larger normal force would also increase the effective compressed area between the surfaces, as such the pressure is distributed over this larger effective compressed area. The relationship between the normal force and the common angle is therefore non-linear. Overall, static friction is shown to depend on the normal force, apparent contact area, Young's modulus, and the compressed efficiency ratio (effective compressed area per apparent contact area). Experimental measurements using teflon were carried out, and the results confirm predictions of this theory.
\end{abstract}


\maketitle


\section{Introduction}

Resistance to motion is a phenomenon that has long been observed by mankind. Aristotle once taught that the force on an object is proportional to its velocity \cite{Arist}. This paradigm of Aristotle on the requirement of a force to maintain an object's velocity is equivalent to Newton's view that an external force is needed to overcome friction. As much as the effects of friction like damping and energy lost in a system could arise from Aristotle's law of motion, such a law is however, non-reversible due to its velocity dependence. After an object comes to a halt due to friction \footnote{Kinetic friction acts when the two surfaces are in relative motion, and static friction takes over (to overcome any net external force) when the relative motion is zero.}, it is impossible to tell which direction it came from \footnote{This is clearly depicted by its phase space, where the trajectories of such a system does not have a conserved volume since they converge to the point of zero momentum (or velocity), violating Liouville's theorem. Indeed, Aristotle's force law cannot be derived from an action principle as it is non-conservative, and it cannot be expressed as the gradient of some potential.}. A non-reversible theory of friction cannot be fundamental, since classical mechanics tells us that classical laws of motion are reversible if all the degrees of freedom of a system are accounted for \cite{Class}. Quantum mechanics is also reversible in the form of its unitary operators which are always invertible \cite{Quant}. It is thus desirable to construct a theory of friction which possesses the feature of reversibility.

Friction is the result of ignoring the astronomical amount of degrees of freedom of the molecules making up a macroscopic object, where following every single one would be intractable. Although a system with many degrees of freedom may be treated by reducing it to a small number of variables, such effective equations always contain damping and the resulting theory would be non-reversible \cite{Fok}. Nevertheless, a simple theory of static friction can be traced back to Amontons \cite{Amon}, stating essentially that the maximum static friction is proportional to normal force, and is independent of area. Over time, there has been plenty of work and effort devoted to this subject due to its immense influence on everyday life, with many excellent references available \cite{Amon1,Amon2,Amon3}. More recent theoretical work and experimental results however, suggest deviations from Amontons' two laws \cite{Exper0,Exper1,Exper2,Exper3,Dev1}, especially by considering various aspects of the microscopic origins of static friction \cite{Amon2,Amon3,Micro1}. In particular, an important notion is that the effective compressed area \footnote{The usual term is ``effective contact area''. In our paper, we would call this ``effective \emph{compressed} area'' because we are considering actual compression between the surfaces in contact that would change this value.} between the surfaces, which is only a very small fraction of the apparent contact area (typically of the order of $10^{-6}$ -- see for example section 4.3.1 in Ref. \cite{Pobell} which mentions this for metals), would increase when a greater load is placed \cite{Arch1,Arch2,Arch3,Arch4,Arch5}. This has helped to explain previous known experimental results which reported departures from Amontons' two laws.

To construct a reversible theory of friction, we model the asperity between the surfaces of contact as a regular array of microscopic inclines and investigate the consequences of such an exact representation for the contact region. This is unlike conventional theories like stick-slip \cite{stickslip0,stickslip1,stickslip2} and the Prandtl-Tomlinson models \cite{tomlin1,tomlin2} which depend on different sets of empirically motivated assumptions, viz. the former studying the jerking motion when two surfaces slide over each other, and the latter based on a point mass driven over a periodic potential. There are also other existing analytical models that are built upon empirical observations and statistical techniques, like Ref. \cite{ana1,ana2,ana3} for example. (See for instance Ref. \cite{ana1} which provides a good summary on other theoretical developments.) Our goal here not to add another such model, but to formulate a theory of friction from fundamental rules of classical mechanics which retains the feature of reversibility, based on the regular microscopic inclines representation.

As a preliminary step towards formulating this theory, we focus solely on the development of static friction in this paper, with the mechanics involving kinetic friction and dynamics to be dealt with subsequently in a future study. To ensure reversibility, we argue that friction does not exist at the microscopic scale, and the macroscopic phenomenon of friction arises as a result of the usual laws of classical physics \footnote{Consider a simple harmonic spring-mass system as an example, whose force is given by Hooke's law and motion governed by Newton's second law. Such a system is reversible, and Liouville's theorem is satisfied. Similarly, a theory of friction based on Newton's second law of motion and an elastic force law would guarantee reversibility.}. In addition, the region of contact between homogeneously polished surfaces is modeled as being composed of a regular array of compressible elastic microscopic inclines. This theory would then be relatively simple and analytically tractable for gleaning useful physical insights, avoiding the need for sophisticated statistical averaging over the asperities. The perception of static friction would be interpreted as the resistance in having to push the load over these smooth microscopic inclines. As the increase in the normal force between the surfaces would compress the microscopic inclines and hence reduce the angle of inclination, we derive a theoretical expression for a non-constant coefficient of static friction. We also take into account that an increase in load would progressively increase the effective compressed area between the two surfaces, so that the pressure is then spread over a larger effective area. This derivation would yield a linear differential equation that must be satisfied to allow the property that experiments carried out on the same material but with different apparent contact areas would obey the same law. From its solution, these would lead to a non-linear relationship between static friction and the load (the normal force between the two surfaces). To test this theory, we carried out experiments for teflon-on-teflon, to find results in agreement with our theory.

This paper is organised as follows: We describe the theory in the next section, with experimental details in Section 3. Section 4 contains a more extensive discussion on the implications of our theory with regards to the experimental results, before we conclude in Section 5.

\section{Our theory: A regular array of compressible elastic smooth microscopic inclines}

\begin{figure}
\includegraphics[width=16cm]{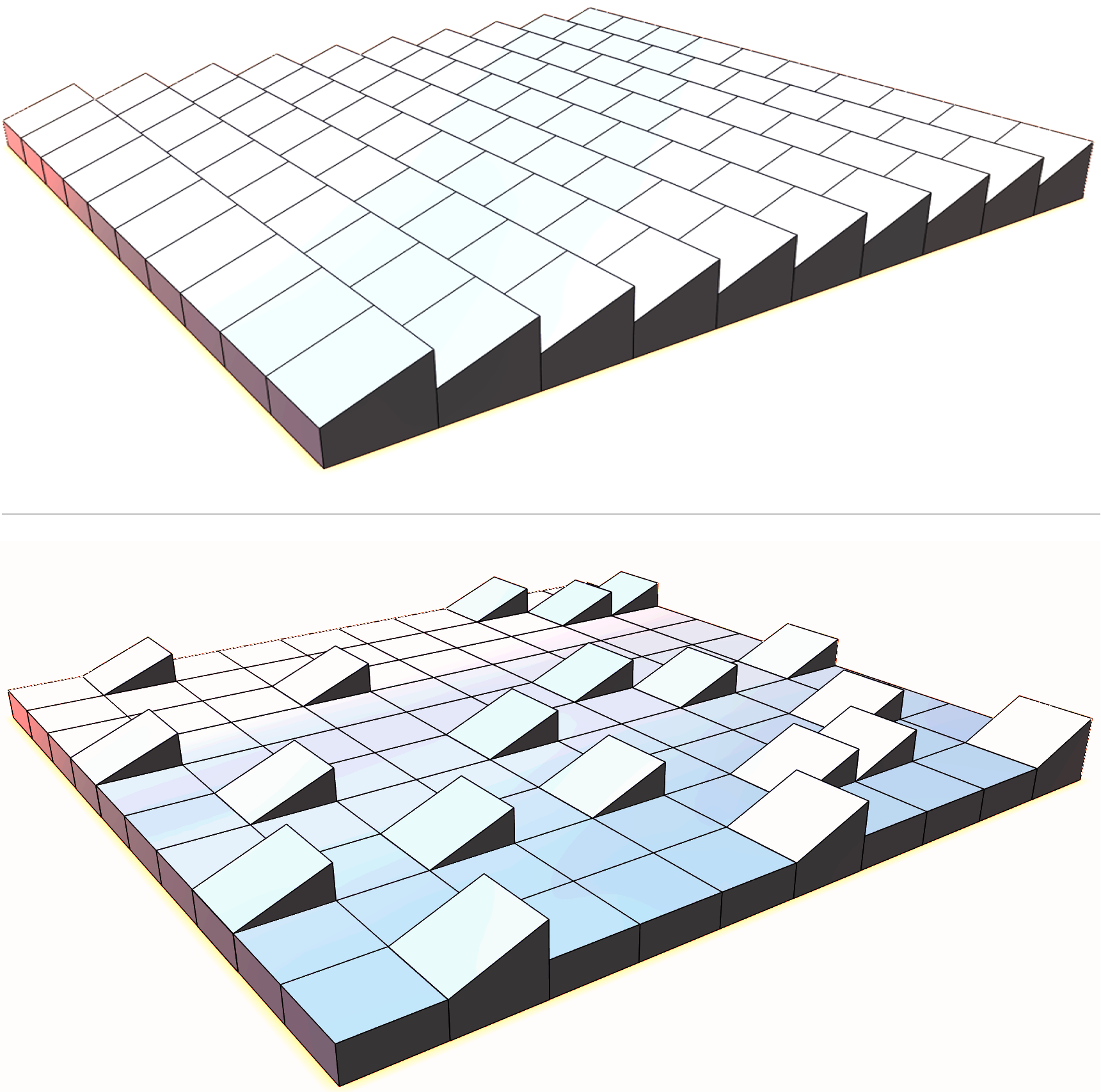}
\caption{The top figure shows how we model the region between two surfaces as a regular array of microscopic inclines with a common angle. The bottom figure takes into account the fact that this region is mostly empty space where the two surfaces are actually not in contact. In this example, there are only 20 microscopic inclines out of a total of 100, so that the compressed efficiency ratio is $\eta=0.2$.}\label{modeling}
\end{figure}

\begin{figure}
\includegraphics[width=16cm]{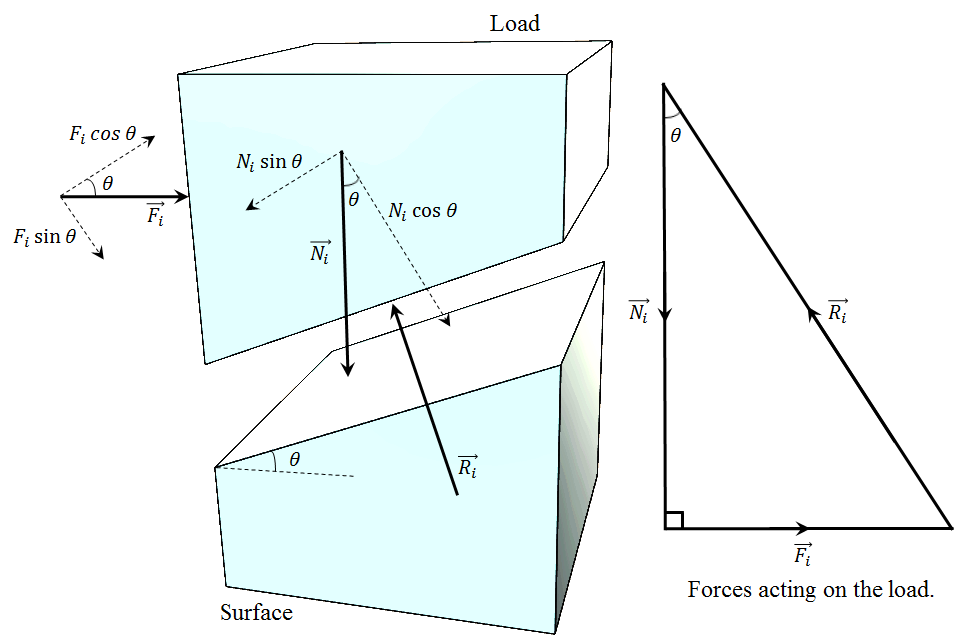}
\caption{Looking at one microscopic incline, representing a region of contact between the asperities of two surfaces: the forces acting on the load and the perceived macroscopic phenomenon of static friction arising from the need to get the load across the microscopic incline. To do so, it is necessary for an external horizontal force $F_i$ to satisfy $F_i\cos{\theta}\geq N_i\sin{\theta}$, where $N_i$ is the weight of the load. The minimum horizontal force would be $F_i=N_i\tan{\theta}$, so that the overall static friction would then be the sum from all these microscopic inclines, $F_s=N\tan{\theta}$. The contact force between the microscopic incline is $R_i$. Note that $N_i=R_i \cos{\theta}$ or $R_i=N_i \sec{\theta}$.}\label{FBD}
\end{figure}

Our theory of static friction is based on the notion that the region of contact between the surfaces is composed of a regular array of compressible elastic microscopic inclines with a common effective angle $\theta$. For regions where there is no contact, there would be no microscopic incline present so that our model becomes an array of regular microscopic inclines where most of these inclines are absent (see Fig. \ref{modeling}). In fact, since it is known that the effective contact area is about $10^{-6}$ of the total apparent area \cite{Pobell}, we would expect in our model that out of a million microscopic inclines, all except one are missing from this regular array. A central idea in our theory is that there is fundamentally no friction at the microscopic scale, i.e. these microscopic inclines are perfectly smooth. The resistance to motion then arises due to the need to push the load over these microscopic inclines, where the perceived resistive force is a component of the normal force when resolved along the direction of the slope (see Fig. \ref{FBD}).

From Newton's second law of motion, the static friction due to climbing the microscopic inclines would be $F_s=N\tan{\theta}$, where $N$ is the usual macroscopic normal force between the surfaces (and $N$ equals the weight of the load when it rests on a horizontal surface). Letting $\mu$ be the coefficient of static friction, i.e. the ratio of $F_s$ to $N$, the coefficient is readily identified as $\mu=\tan{\theta}$. If these microscopic inclines are rigid or non-deformable, then $\theta$ is a constant so that the coefficient is a constant, in accordance with Amontons' law that $F_s$ is proportional to $N$. It is therefore interesting that consideration of the elasticity of the microscopic inclines would lead to a more extensive description of static friction. Upon its complete formulation, we show that Amontons' two laws appear as the \emph{rigid object limit} of our theory (see end of this section).

\begin{figure}
\includegraphics[width=8cm]{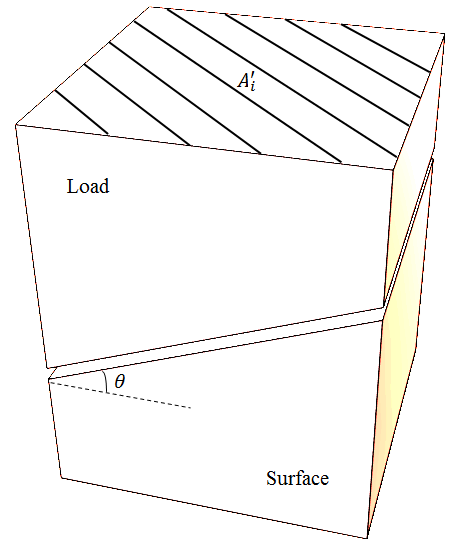}
\caption{The effective compressed area $A'_i$ of a microscopic incline is defined as the horizontal projection of the area it occupies. If $\bar{A}_i$ is the area of the slanted plane where the load and the surface are in contact, then $\bar{A}_i\cos{\theta}=A'_i$ or $\bar{A}_i=A'_i\sec{\theta}$. The total $A'$ is the sum due to all such microscopic inclines, $A'=\sum{A'_i}$.}\label{area}
\end{figure}

Real materials are non-rigid, and a measure of the elastic stress due to compression is given by Young's modulus $E$. The normal force would elastically compress the microscopic inclines from its uncompressed angle $\theta_0$ to a smaller angle $\theta$, so that $\mu=\tan{\theta}$ would decrease. This stress can be related to the angular Young's modulus $E'$ and angular strain, viz. stress equals $E'$ times angular strain, in analogy with stress equals $E$ times linear strain. The angular Young's modulus is not equal to the Young's modulus but is slightly larger because it is somewhat harder to compress through an angle (due to resistance from behind the wedge) than to directly compress normally, so we posit that $E'=\chi E$ where $\chi>1$ is a constant. The pressure due to the normal force $N$ is only distributed over the effective compressed area $A'$, where $A'=\eta A$ (see Fig. \ref{area}). Here, $A$ is the apparent contact area, i.e. the macroscopically measured area of the load that is purportedly in contact with the surface. We show later that the compressed efficiency ratio, $\eta\in[0,1]$ would increase with increasing $N$.

We can therefore write the stress-angular strain relation as
\begin{eqnarray}
\frac{R_i}{\bar{A}_i}=\frac{N_i\sec{\theta}}{A'_i\sec{\theta}}=\frac{N_i}{A'_i}=\frac{N}{A'}=\frac{N}{\eta A}=\chi E\left(\frac{\theta_0-\theta}{\theta_0}\right),
\end{eqnarray}
using the fact that the stress due to one microscopic incline $N_i/A'_i$ (recall that the subscript $i$ represents one single microscopic incline) is equivalent to the macroscopic stress of the entire load $N/A'$ (where the absence of the subscript $i$ denotes the macroscopic quantities, i.e. the sum over all microscopic inclines). The resulting equation can be rewritten as
\begin{eqnarray}\label{second}
\theta=\theta_0\left(1-\frac{N}{\chi\eta EA}\right).
\end{eqnarray}
This illustrates that if $\eta$ is constant (or if its change is fairly negligible), then $\theta$ and $N$ have a linear relationship.

From the linear stress-strain relation $N/A=E\varepsilon$ where $\varepsilon=N/EA$ is the linear strain \footnote{For this linear stress-strain relation, we are adopting the macroscopic picture where the load sits on the apparent contact area $A$, so that $\varepsilon=N/EA$ does not depend on $\eta$.}, we can also write the above equation as
\begin{eqnarray}\label{strain}
\theta=\theta_0\left(1-\frac{\varepsilon}{\chi\eta}\right).
\end{eqnarray}
Although we would be measuring the variation of static friction due to different loads and apparent contact areas, it is instructive to alternatively express the quantities in terms of linear strain (Eq. (\ref{strain})) as it is a dimensionless quantity, representing the amount of compression per original length. Note that $\chi$ and $\eta$ are also dimensionless.

As the load increases, $\eta$ would monotonically increase due to the increasing stress and asymptotically approach a limit $\eta_\infty$ which may not necessarily be 1, especially if the two surfaces can never be perfectly squashed without having any more gaps in between. It is convenient to express the change in $\eta$ with respect to $\varepsilon$ as a Taylor expansion:
\begin{eqnarray}\label{etagen}
\frac{d\eta}{d\varepsilon}=k_1(\eta_{\infty}-\eta)+k_2(\eta_{\infty}-\eta)^2+\cdots,
\end{eqnarray}
where $k_i$'s are constants. It is clear that a differential equation of such a form has the properties that $\eta$ increases monotonically with $\varepsilon$, and the rate of increase would diminish as $\eta\rightarrow\eta_\infty$.

The behaviour of $\eta$ should be independent of whether the strain is due to a load with large or small apparent contact areas, since one is free to conduct an experiment with any desired dimensions. In particular, if $\eta_A$ for a load of apparent area $A_A$ and $\eta_B$ for a load of apparent area $A_B$ are solutions to Eq. (\ref{etagen}), then $\eta_{AB}$ for a load of apparent area $A_A+A_B$ would also be a solution to the same differential equation. This linear combination of solutions being a solution itself would be guaranteed if that said differential equation (Eq. (\ref{etagen})) is linear \footnote{An analogy would be the requirement of the wave equation being linear, to readily imply that the linear superposition of two waves (satisfying the wave equation) is also a wave (which satisfies the wave equation as well).}. Hence the form of the equation for $\eta$ would be
\begin{eqnarray}\label{eta}
\frac{d\eta}{d\varepsilon}=k(\eta_{\infty}-\eta),
\end{eqnarray}
where $k_1$ is renamed to $k$ since it is the only non-zero constant. To further clarify this point, an explicit proof of the following is given in the appendix: The independence of the form of the differential equation governing $\eta$ with regards to the apparent contact area implies that it is a linear differential equation.

Solving Eq. (\ref{eta}) gives
\begin{eqnarray}
\eta=(\eta_\infty-\eta_0)(1-e^{-k\varepsilon})+\eta_0,
\end{eqnarray}
where $\eta_0$ is the compressed efficiency ratio under no load (or no strain). Note that $k$ is also a dimensionless quantity, alongside $\eta$ ($\eta_0$, $\eta_\infty$) and $\chi$. We might expect $k$ to be of order 1, due to our dimensionless formulation of $\eta\in[0,1]$ as a function of $\varepsilon\in[0,1]$. Later in Section 4, we find that curve fitting onto the experimental data indeed gives $k$ fairly close to 1.

We can ultimately put all these together to give
\begin{eqnarray}
F_s=N\tan{\left(\theta_0\left(1-\frac{N}{\chi\eta EA}\right)\right)},
\end{eqnarray}
where $\eta=(\eta_\infty-\eta_0)(1-e^{-k\varepsilon})+\eta_0$ and keeping in mind that $\varepsilon=N/EA$. It turns out that it is useful to consider the product of $\chi$ and $\eta$ as a whole when comparing with experimental measurements (see Section 4), so let $\zeta=\chi\eta=(\zeta_\infty-\zeta_0)(1-e^{-k\varepsilon})+\zeta_0$, where $\zeta_0=\chi\eta_0$ and $\zeta_\infty=\chi\eta_\infty$. The above equation would then become
\begin{eqnarray}\label{ultim}
F_s=N\tan{\left(\theta_0\left(1-\frac{N}{\zeta EA}\right)\right)}=N\tan{\left(\theta_0\left(1-\frac{\varepsilon}{\zeta}\right)\right)}.
\end{eqnarray}
An important observation from Eq. (\ref{strain}) is that $\theta=0$ when $\varepsilon=1$ since a full angular compression corresponds to the microscopic inclines having zero inclination (so that it is perfectly flat). This imposes a boundary condition: $\zeta=1$ when $\varepsilon=1$ \footnote{Despite being flat when $\varepsilon=1$, it may not be true that $\eta_\infty=1$ since the irregularity of the asperities may result in some parts of the two surfaces always remaining untouched.} or equivalently,
\begin{eqnarray}\label{boundary}
\zeta_\infty=\chi\eta_\infty=\frac{1-\zeta_0 e^{-k}}{1-e^{-k}}>1.
\end{eqnarray}
Since $\zeta_0=\chi\eta_0\sim10^{-6}$ and $k\sim1$ (later, we would find that our experimental results do indeed yield such orders for $\zeta_0$ and $k$, as summarised in Table 1), it may be useful to approximate $\zeta_\infty\approx(1-e^{-k})^{-1}$ for practical purposes \footnote{In fixing the boundary condition for $\zeta_\infty$, we are of course assuming that stress is proportional to strain where the Young's modulus is always a constant. Under extreme stress beyond the proportionality limit or past the elastic limit, this would not be true. Nevertheless in this theory, we consider the simplest relation ``stress equals Young's modulus times strain'' since the range of our experimental measurements remains well within the proportionality limit, and this boundary condition suffices.}.

With our theory now complete, let us examine how Amontons' laws are contained within it. For a very small range of $N$, it may be assumed that $\zeta$ (or $\eta$) remains constant so that the argument of the tangent in Eq. (\ref{ultim}) decreases linearly with increasing $N$ (for some fixed $A$). Moreover, if $N$ is small compared to $\zeta EA$, then it leaves this argument as the constant $\theta_0$, reproducing Amontons' law for static friction: $F_s=\mu N$, with $\mu$ being a constant \footnote{Taking $\zeta=10^{-6}$ with a load of 5 kg on a 15 cm by 15 cm apparent contact area, then $N/\zeta EA$ is of order $10^{-2}$ for metals ($E\approx100$ GPa), but is about 4 for teflon ($E=0.5$ GPa).}. Furthermore, we can also deduce that a change in $A$ would have little effect if $N$ is small compared to $\zeta EA$ since $N/\zeta EA$ would be inconsequential compared to 1 anyway for these different values of $A$, yielding Amontons' other law that $F_s$ is independent of area. To sum it up, the \emph{rigid object limit}, $E\rightarrow\infty$ of our model reduces to Amontons' laws since any finite $N$ and $A$ would then be negligible compared to an infinitely large $E$.

To be able to test the predictions of our theory with regards to the non-Amontons features, we would need to use a material with relatively small $E$, and collect experimental data for various $N$ and $A$. We proceeded to do this for teflon, and the experimental details are given in the next section.

\section{Two independent experiments that measure the maximum static friction between teflon-on-teflon for different loads and apparent contact areas}

Teflon plates of various sizes were machine-polished to attain homogeneity of a particular surface roughness, with different amount of weights then loaded on them. These loaded teflon plates were placed on a similarly polished teflon track. Although ideally we would like to measure from the same teflon plate throughout the experiment (because using different ones would introduce errors due to inevitable variation in homogeneity), this is not possible since we wish to collect data for different $A$ as well. For now we assume that these different plates have an equivalent degree of homogeneity, and later provide an estimated percentage error that different surface roughness produced by the polishing procedure may yield.

Two different experimental techniques were designed and performed independently to determine the maximum static friction between teflon-on-teflon for different $N$ and $A$:
\begin{enumerate}
\item
The first was an inclined plane setup \footnote{The inclined setup experiment was developed by Sunku Sai Swaroop and Vee-Liem Saw, with the latter carrying out the experiment to collect a total of 55 data points for various $N$ and $A$.}, where a loaded teflon plate was placed on a teflon track. One end of the track was raised very slowly via a mounted screw, until the loaded teflon plate started to slide (see Fig. \ref{incsetup}). The height of inclination when this happened was measured using a height gauge, allowing us to calculate the angle of inclination and the maximum static friction between the two teflon surfaces. This experiment was done with good precision, since the screw was turned extremely gradually, and each measurement was repeated four additional times to obtain the average. As many as thirteen to fifteen different loads from around 50 g to 10 kg were used for a particular size of the teflon plate to yield sufficient information for testing our theoretical predictions. This experiment was done for four different values of $A$ by using four sizes of teflon plates: $A$, $A/2$, $A/4$, $A/8$, where $A=0.045$ m$^2$ (Fig. \ref{plates1}). We should remark that in this experimental design, when a load of weight $W$ was placed on the track and the latter tilted such that the former started to slide when at angle $\theta$, the normal force between the surfaces is not $N=W$ but $N=W\cos{\theta}$.

\begin{figure}
\includegraphics[width=16cm]{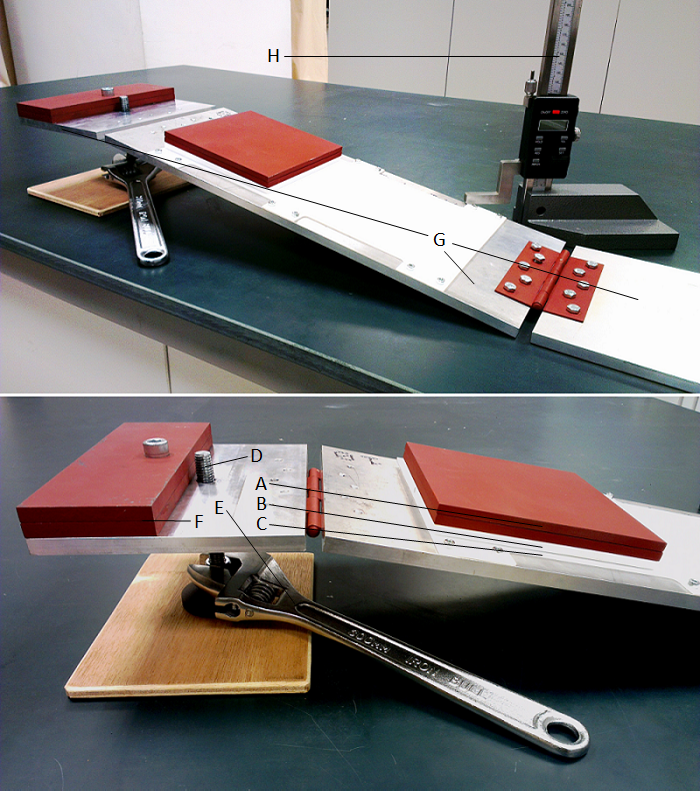}
\caption{Top: The inclined setup, where a loaded teflon plate is placed on an inclined teflon track. Bottom: A close-up view of the screw which is slowly turned with a wrench, to gradually increase the inclination angle of the track. Labeled parts: A) steel load on teflon plate, B) teflon plate sitting on inclined teflon track, C) teflon track supported by an aluminium base, D) the screw which raises one end of the aluminium base, E) wrench for turning the screw, F) counterweight, G) aluminium supports, H) height gauge which measures vertical distances between two points.}\label{incsetup}
\end{figure}

\begin{figure}
\includegraphics[width=13cm]{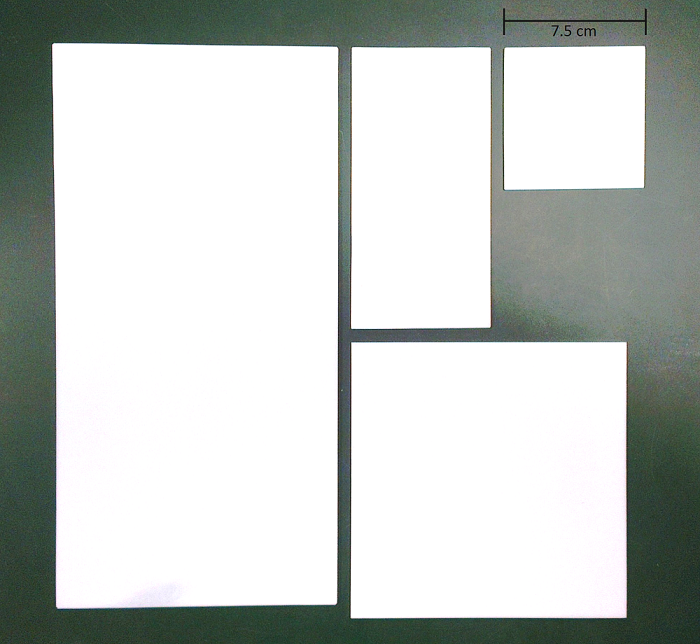}
\caption{The inclined setup measures for the four teflon plates with areas $A$, $A/2$, $A/4$, $A/8$, where $A=30$ cm by 15 cm or 0.045 m$^2$.}\label{plates1}
\end{figure}

\begin{figure}
\includegraphics[width=15cm]{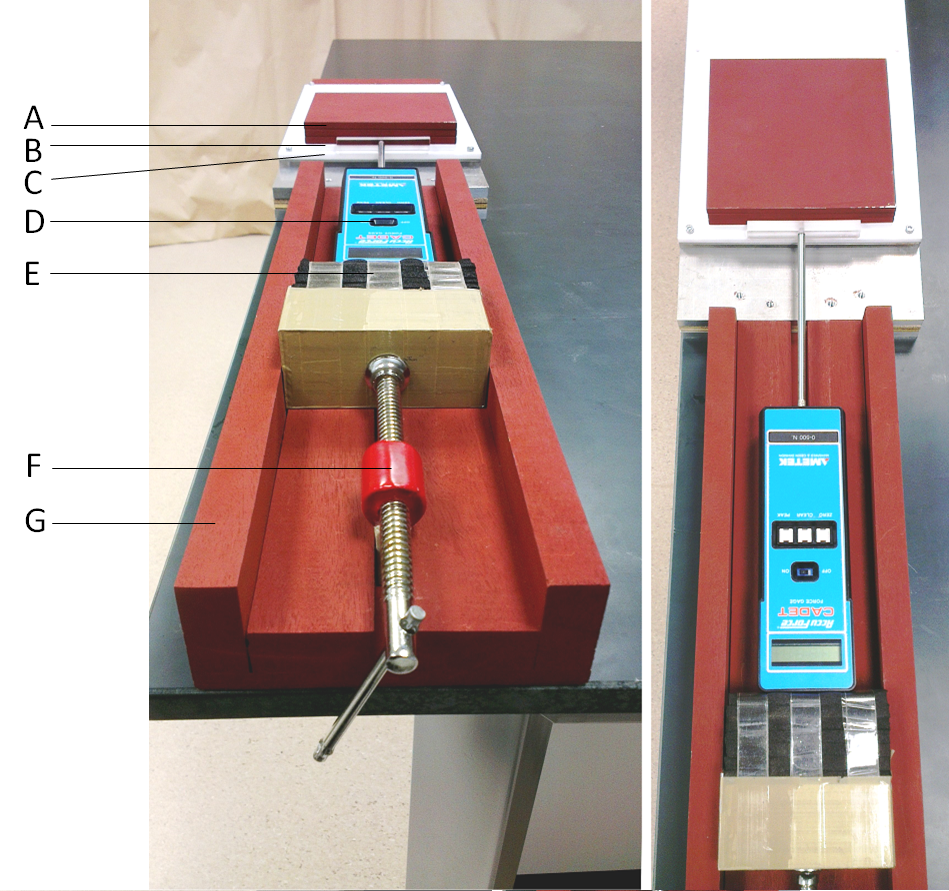}
\caption{The horizontal setup, where a loaded teflon plate is placed on a horizontal teflon track. Labeled parts: A) steel load on teflon plate, B) teflon plate sitting on horizontal teflon track, C) teflon track supported by an aluminium base, D) force gauge, E) absorber, F) a screw which gradually pushes the force gauge, mounted on the supporting wooden frame, G) wooden frame.}\label{horsetup}
\end{figure}

\begin{figure}
\includegraphics[width=16cm]{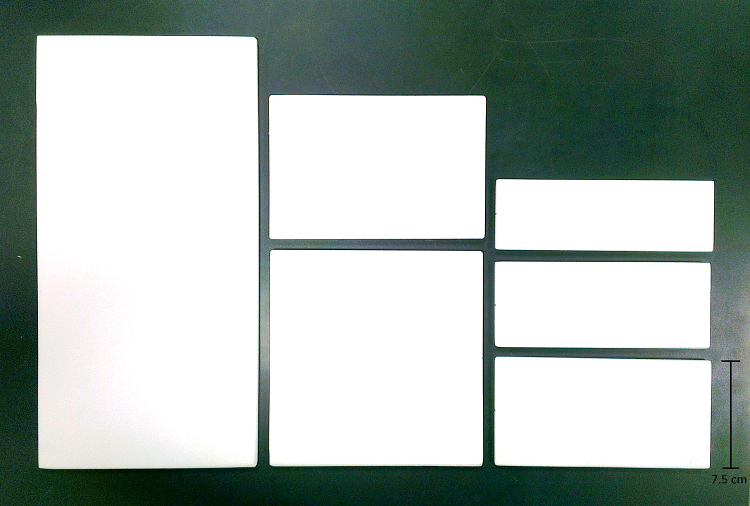}
\caption{The horizontal setup measures for the six teflon plates with areas $A$, $A/2$, $A/3$, $A/4$, $A/5$, $A/6$, where $A=30$ cm by 15 cm or 0.045 m$^2$.}\label{plates2}
\end{figure}

\item
A second experimental method was subsequently carried out \footnote{After the first experiment was completed, the horizontal setup experiment was subsequently designed by Freeman Thun and Kin Sung Chan, and they both collected 36 data points for various $N$ and $A$.}, with the aim of collecting data for six different apparent contact areas. By using a force gauge to push on the load sitting horizontally, the minimum force needed to cause relative motion between the loaded teflon plate and the teflon track was directly measured (see Fig. \ref{horsetup}). Similar to the first experiment, each measurement was taken five times to obtain the average. Six different loads were used for each of the following six areas: $A$, $A/2$, $A/3$, $A/4$, $A/5$, $A/6$, where $A=0.045$ m$^2$ (Fig. \ref{plates2}). Since the track is always horizontal, $N=W$.
\end{enumerate}

These two independent experimental methodology would allow for comparisons between each set of results. The data obtained show that both experiments support our theory (see the discussion in the next section). We chose teflon because it is one of the softest materials (Young's modulus of 0.5 GPa, page 8 of \cite{tefmet}) where metals on the other hand are typically three orders of magnitude harder (Young's modulus in the region of 100 GPa \cite{tefmet}). This choice would allow relatively easier experimental detection of the non-constant feature of static friction coefficient. Although rubber is even softer than teflon, the former is prone to other kinds of frictional causes, viz. adhesion, dragging and shear forces. Attempts were carried out on wood and acrylic, but were met with several difficulties since other sources of friction due to gripping and suction caused by air pressure are non-negligible. A simple test could be conducted to show the existence of suction: when two such surfaces are separated along the normal direction, it was observed that this suction force increases with the separation speed -- which is the opposite of adhesive force that breaks more easily when the separation speed is increased.

Teflon on the other hand does not seem to display much of these effects, making it a reasonably good candidate material to demonstrate the behaviour predicted by our model whereby the resistance to motion arises chiefly due to the geometrical topography of the surfaces as opposed to the intermolecular forces. It is also a material with one of the lowest coefficient of static friction, as low as 0.04 if polished extremely smoothly \cite{Giancoli}. In our experiments, our teflon surfaces were not polished with the finest grain since we only require the surfaces to be homogeneous, regardless of the surface roughness. We thus expect the coefficient of static friction to be higher than 0.04. To examine the finer structure of our polished teflon surfaces, a sample was placed under an optical microscope. In addition, atomic force microscopy (AFM) was carried out to further reveal its 3-d microscopic structure, as presented below.

\begin{figure}
\includegraphics[width=16cm]{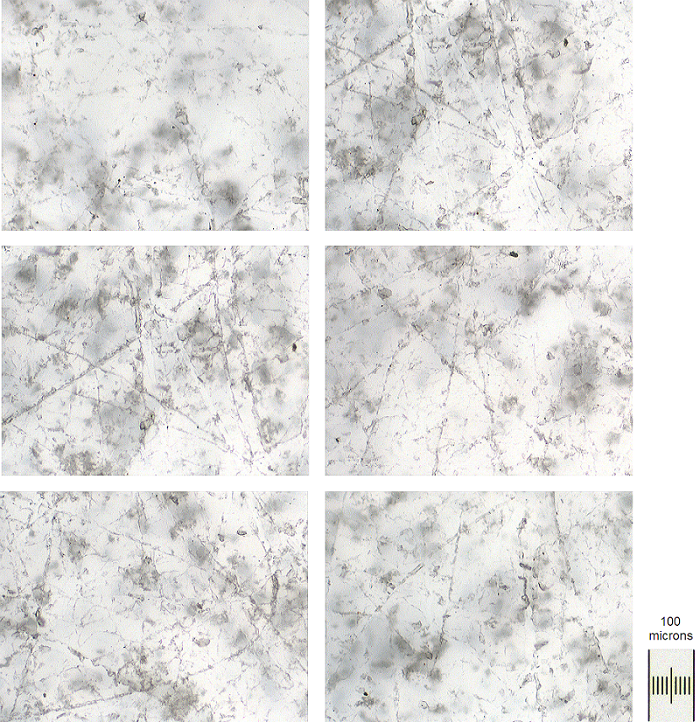}
\caption{Six images from random positions of one of our polished teflons viewed under an optical microscope with $10\times$ magnification.}\label{10x}
\end{figure}

\subsection{Images of homogeneously polished teflon surface at the microscopic level}

\begin{figure}
\includegraphics[width=16cm]{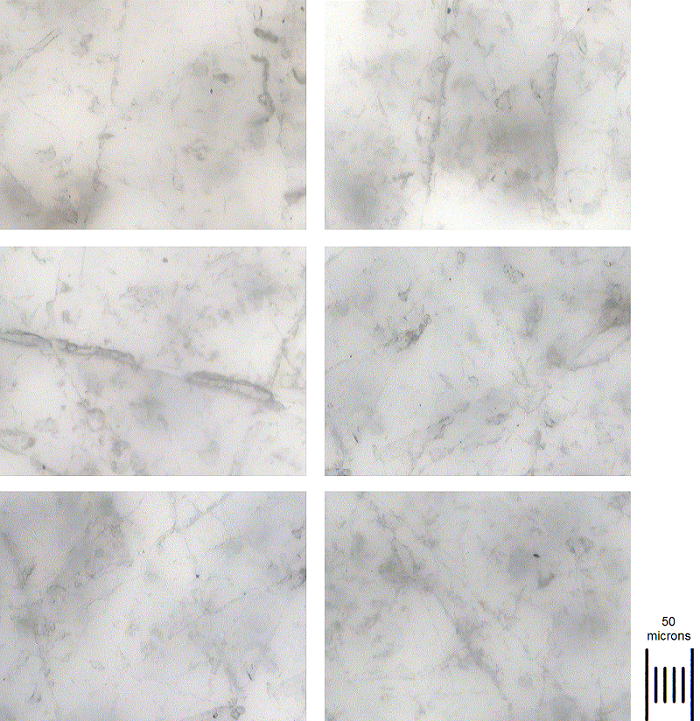}
\caption{Images from six random positions of a polished teflon with $20\times$ magnification.}\label{20x}
\end{figure}

\begin{figure}
\includegraphics[width=16cm]{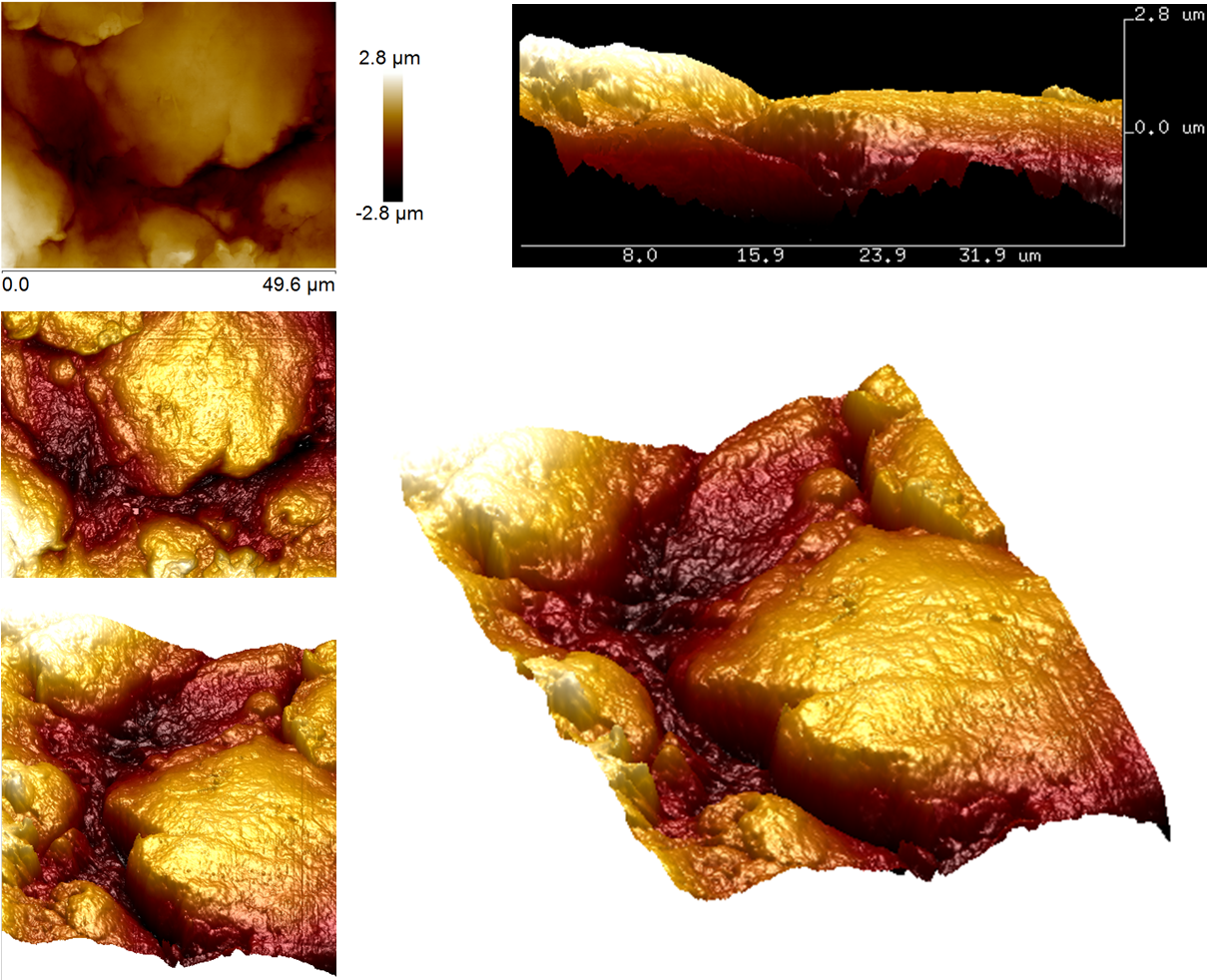}
\caption{An AFM image of one of our polished teflons. Shown here are several viewing angles.}\label{AFM1}
\end{figure}

\begin{figure}
\includegraphics[width=16cm]{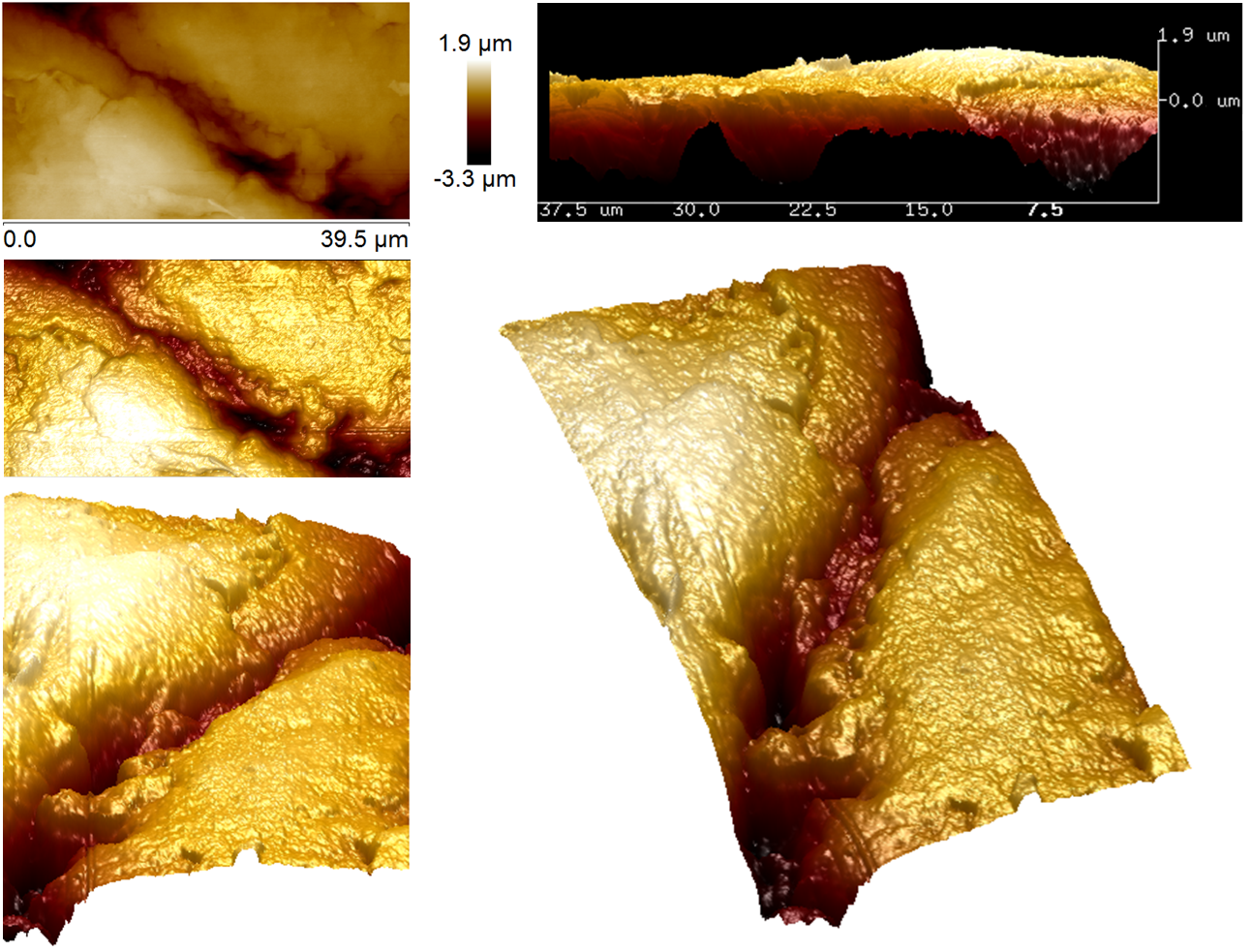}
\caption{A second AFM image at a different position. Shown here are several viewing angles.}\label{AFM2}
\end{figure}

A sample of our machine-polished teflon surfaces was placed under an optical microscope for examination of its micro-structure. The teflon surface is slightly translucent, allowing light from the standard microscope light source to partially go through and hence no additional lighting was required. Images of the teflon surface under $10\times$ and $20\times$ magnifications at random locations were captured, as shown in Figs. \ref{10x} and \ref{20x} respectively. The images under $10\times$ magnification are relatively sharp, although they are accompanied by shades of smear. The images under $20\times$ magnification are not as sharp, because it was harder to focus. We also tried greater magnification but found that it was increasingly challenging to focus and produce clear images. The appearances of the smears (as can be seen in Figs. \ref{10x} and \ref{20x}) due to the difficulty in focusing is caused by the uneven surface with varying depth of several microns. In spite of the limited use of the optical microscope which does not really illustrate the 3-d structure, it nevertheless provides a good indicative picture of the surface terrain around the order of hundreds of microns which would not be obvious to the naked eye. To probe even further towards a finer scale of one to tens of microns (with sub-micron resolution) and glimpse its 3-d microscopic structure, AFM images were obtained and shown in Figs. \ref{AFM1} and \ref{AFM2}.

The images show that the polished teflon surface is not a perfectly smooth flat plane. Instead, it is more of ragged badlands with couloir-like topography as is vividly depicted by Figs. \ref{10x}--\ref{AFM2}. These gully structures are actually vaguely visible to the naked eye if one painstakingly stares at it, although the purely white appearance of teflon makes it difficult to pick out these features due to lack of contrast. Such gorges are typically of 5-30 microns wide (consistently found in the images under $10\times$ and $20\times$ magnifications as well as the AFM ones) and several microns in depth, with some possibly stretching across the entire length of the teflon surface (see the $10\times$ magnification images in Fig. \ref{10x}).

Although not shown here, an unpolished teflon surface would naturally have rough or non-flat apexes, which would lead to greater friction between two such surfaces. After polishing, the sharper peaks are chopped off into relatively flat plateaus. The AFM scans confirm that the polished teflon surface comprises slightly inclined or rounded off mesas, separated by the ravines. It may be worth noting that these mesas typically span about several tens of microns wide with height difference (from lowest point to highest point of a mesa) of a couple of microns, giving an estimated inclination angle of roughly between 0.01 rad to 0.2 rad. With such micro-structure, it is therefore reasonable to represent the contact between two teflon surfaces by our model of an array of regular microscopic inclines where most of the inclines are absent as in Fig. \ref{modeling}, since two such surfaces would actually only be in contact over small areas. 

\section{Results and discussion}

\begin{figure}
\includegraphics[width=15cm]{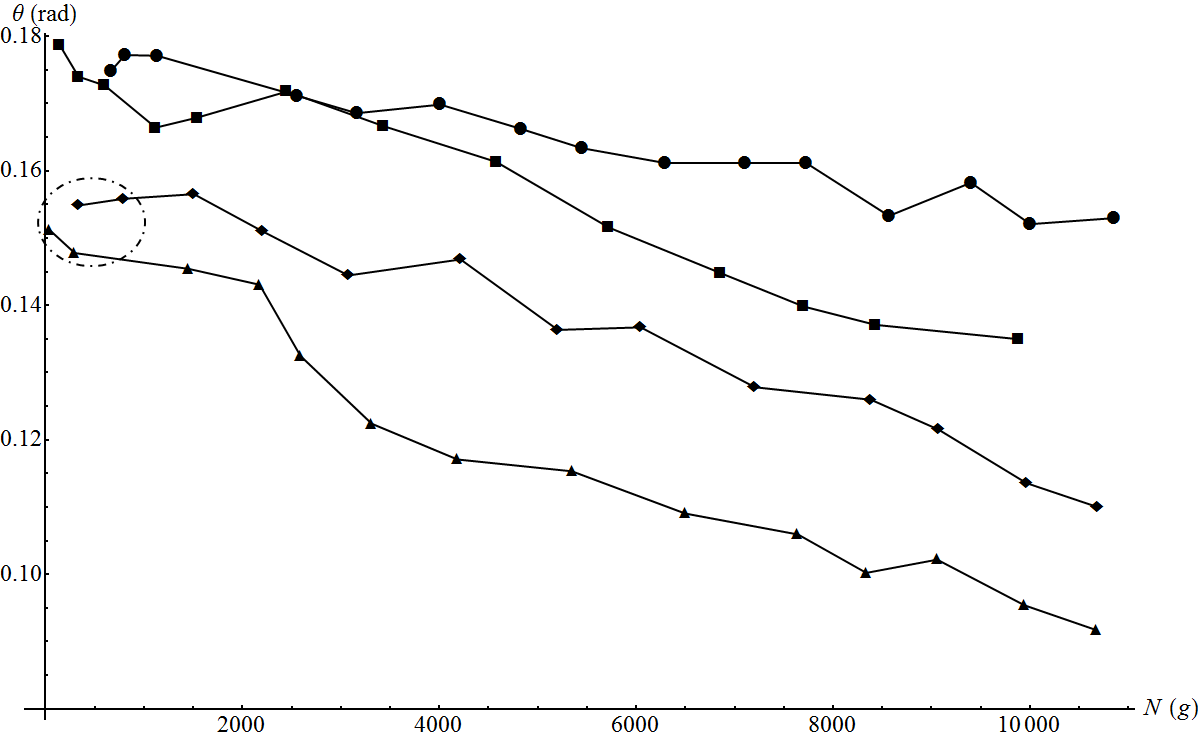}
\caption{Plots of $\theta$ against $N$ for $A$ (filled circle), $A/2$ (filled square), $A/4$ (filled diamond), $A/8$ (filled triangle), where $A=30$ cm by 15 cm or 0.045 m$^2$. The error bars are not included, as they are small. (Comments on the circled points are given later, in the discussion on Fig. \ref{thetaN2}.)}
\label{thetaN}
\end{figure}

Fig. \ref{thetaN} shows the angle $\theta$ as a function of the normal force $N$ for the first experiment (inclined setup), for four different areas $A$, $A/2$, $A/4$, $A/8$, where $A=30$ cm by 15 cm or 0.045 m$^2$. Note that the actual angle of inclination of the teflon track (which is calculated from our measurement of the height inclination) is the same as the angle $\theta$ of the microscopic inclines from our model. Also, the normal force is related to the weight of the load $W$ by $N=W\cos{\theta}$. These plots show that the coefficient ($\tan{\theta}$) generally decreases with increasing $N$, with the drop more significant for smaller area. It can be seen that the gradient of the plot for $A/2$ (filled square) appears to be about twice as steep as that of $A$ (filled circle), so that the compressed efficiency ratio $\eta$ is fairly constant. However, the gradient of the plot for $A/8$ (filled triangle) is not twice as steep as that of $A/4$ (filled diamond), which implies that the increase in $\eta$ (due to greater stress on $A/8$) cannot be ignored. From Eq. (\ref{second}), we see that a larger $\eta$ has the effect of lowering the gradient of the $\theta$ against $N$ line.

\begin{figure}
\includegraphics[width=15cm]{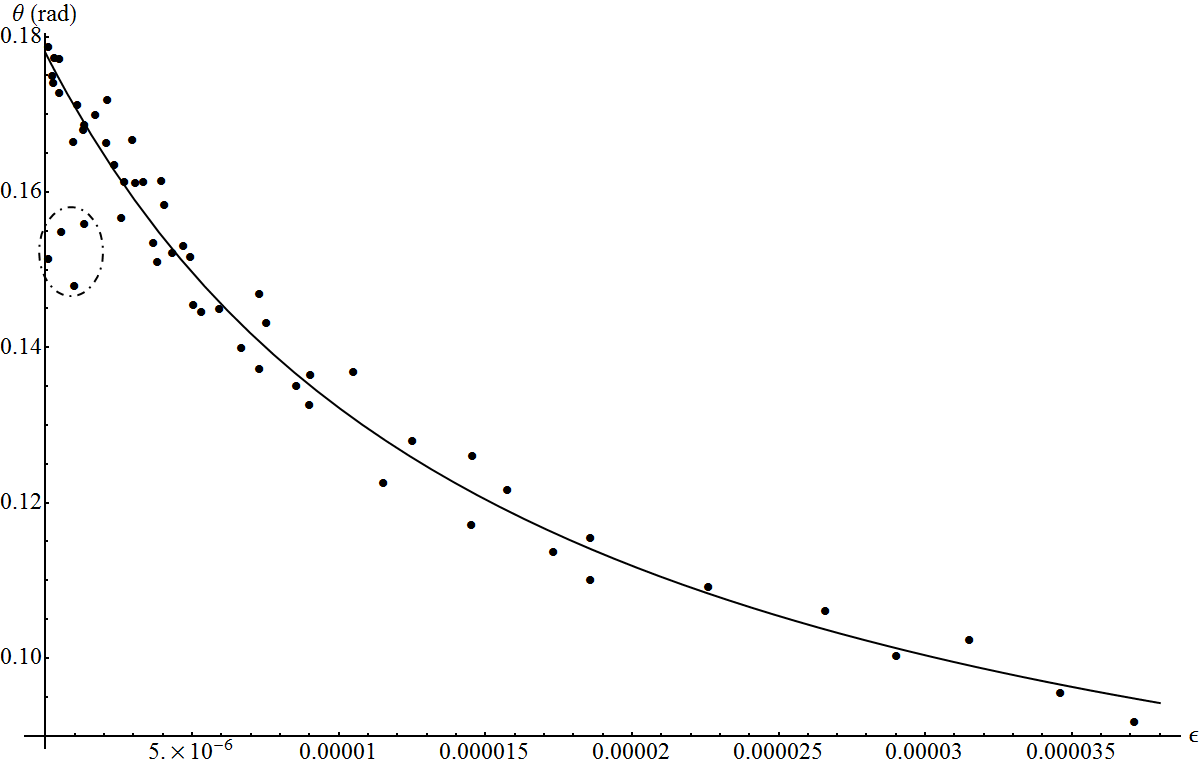}
\caption{Graph of $\theta$ against $\varepsilon$ for various loads and apparent contact areas in the first experiment, combining the data points from all samples. Curve fitting the data to Eq. (\ref{fit}) gives $\zeta_0=(2.40\pm0.18)\times10^{-5}$ and $k=0.86\pm0.17$. The adjusted $r^2$ value for the curve fitting is 0.923. See footnote [42] for a remark on the circled four points.}
\label{thetaN2}
\end{figure}

We can combine all these data points into an overall graph of $\theta$ against $\varepsilon$ (the linear strain) to be plotted and fitted with the theoretical curve given by
\begin{eqnarray}\label{fit}
\theta=\theta_0\left(1-\frac{\varepsilon}{(\zeta_\infty-\zeta_0)(1-e^{-k\varepsilon})+\zeta_0}\right).
\end{eqnarray}
Before this curve fitting is carried out, the value of $\theta_0=(0.178\pm0.001)$ radians is found by fitting a straight line to the fifteen data points for the teflon plate of size $A$ (the largest one). The range of the strain for this is relatively small so that we may assume that $\zeta$ is fairly unchanging and therefore $\theta$ decreases linearly with $N$. With this, there are only two free parameters $\zeta_0$ and $k$ in Eq. (\ref{fit}) (recall that $\zeta_\infty$ is given by Eq. (\ref{boundary})). The curve fitting gives $\zeta_0=(2.40\pm0.18)\times10^{-5}$ and $k=0.86\pm0.17$, with the graph shown in Fig. \ref{thetaN2} (adjusted $r^2$ value of 0.923 \footnote{It is intriguing to remark that if those four points inside the dotted region of Fig. \ref{thetaN2} are ignored and Eq. (\ref{fit}) is then fitted with the three free parameters $\theta_0$, $\zeta_0$ and $k$, the resulting fitted parameters would be $\theta_0=(0.179\pm0.001)$ rad, $\zeta_0=(2.43\pm0.17)\times10^{-5}$, $k=0.81\pm0.11$, with the adjusted $r^2$ value of 0.980. Note that these four points in Fig. \ref{thetaN} correspond to small loads on the teflon plates with areas $A/4$ and $A/8$. The deviations of these four points suggest that the teflon surfaces might have undergone permanent deformation (which would slightly alter the surface roughness) due to the extreme stress of the heavier loads over these smaller areas.}). The value of $\zeta_\infty$ can then be calculated from Eq. (\ref{boundary}) to be $1.73\pm0.22$.

The error bars are not included, as they are small. The main cause of the scatter in the data points is that in spite of our best efforts in polishing the plates to attain the highest possible degree of homogeneity, measurements taken over different plates would inevitably generate random errors. To estimate the error due to sample-to-sample variations, we collected measurements for a particular value of strain: $\varepsilon=1.30\times10^{-6}$, for four different plates (since this experiment uses four plates) of varying sizes and loads (to give this value of $\varepsilon$). The mean value of the corresponding angle is $\theta_{\textrm{mean}}=0.166$ rad with a standard deviation of 0.007 rad, indicating that if one were to carry out this particular experiment to measure the strain by using different but similarly polished plates, there would be an uncertainty of $4.1 \%$ or percentage error per sample of $1.0 \%$.

The general trend in Fig. \ref{thetaN2} is clear -- it nicely follows our model, with $k$ being very close to 1. There are more points per unit strain towards the smaller strain values as compared to the larger ones in the $\theta$ vs $\varepsilon$ graph, due to the conversion of $N$ and $A$ for each of the four different areas to $\varepsilon$: a range of $N$ from 100 g to 10 kg on size $A$ is equivalent to a range of 12.5 g to 1.25 kg on size $A/8$, etc. \footnote{Similarly, a range of $N$ from 100 g to 10 kg on size $A/2$ is equivalent to a range of 25 g to 2.5 kg on size $A/8$, and a range of $N$ from 100 g to 10 kg on size $A/4$ is equivalent to a range of 50 g to 5 kg on size $A/8$.}. In fact, the effective overall range of this whole experiment is essentially tantamount to measuring a range of $N$ from 100 g to nearly 100 kg on an area of $A=0.045$ m$^2$, i.e. about three orders of magnitude.

As mentioned in Section 2, the value of $\eta_\infty$ is not necessarily equal to 1 due to the irregularity of the asperities, especially if there are still gaps when the microscopic inclines are fully compressed. By the assumption of homogeneity, for every microscopic incline oriented in a certain direction, there should be one in the opposite direction. The one in the opposite direction offers no resistance and thus can be treated as if there is none present. This implies that $\eta_\infty$ is only 0.5. Moreover, there could be other orientations which further reduce the effective compressed area and thus lower the value of $\eta_\infty$ even more. Supposing that $\eta_\infty=0.2$, we then find that $\chi=8.64\pm1.08$ and $\eta_0=(2.78\pm0.28)\times10^{-6}$. These parameters turn out to be of the right order, since $\chi$ represents how much more difficult it is to have an angular compression versus a direct inward compression, and it is well-known from other previous work that $\eta_0$ is typically of the order of $10^{-6}$ \cite{Pobell}.

\begin{figure}
\includegraphics[width=15cm]{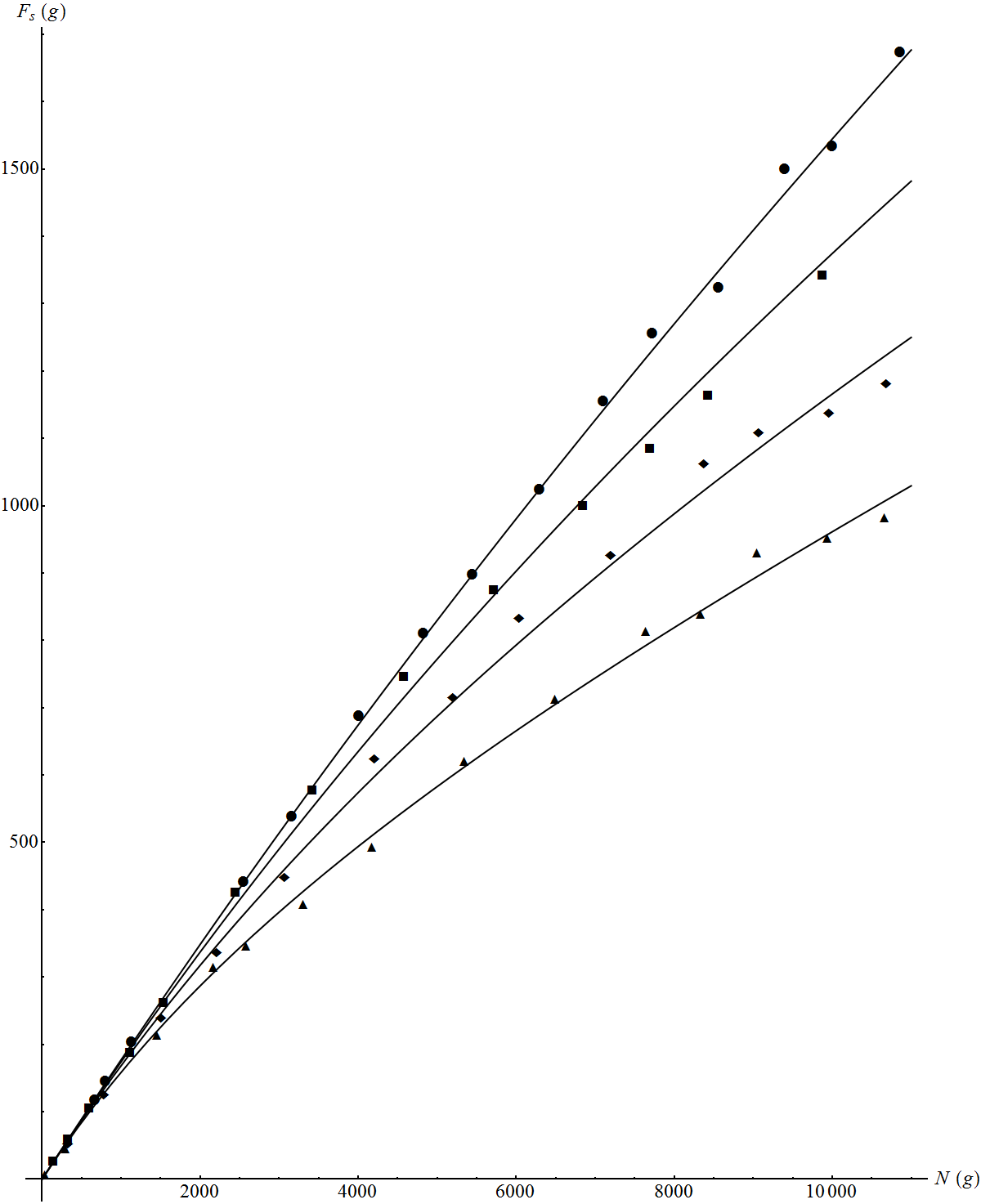}
\caption{Graphs of $F_s$ against $N$ plotted with Eq. (\ref{ultim}) for $A$ (filled circle), $A/2$ (filled square), $A/4$ (filled diamond), $A/8$ (filled triangle). The smaller the area, the greater is the dip of the curve, showing that the coefficient of static friction reduces as the normal force increases, and this is magnified for smaller area (larger pressure).}
\label{FN}
\end{figure}

With the values of the free parameters $\zeta_0$ and $k$ obtained by curve fitting Eq. (\ref{fit}), we can now graph Eq. (\ref{ultim}) onto the plot of maximum static friction against normal force for each of the four areas in Fig. \ref{FN}, illustrating the non-linearity which implies a non-constant coefficient, contrary to Amontons' law. Note that the dip of the curve is more prominent for smaller areas, which correspond to greater stress (force per unit area). This area dependence is obvious for teflon since its Young's modulus is about three orders smaller than that of metals, which explains the lack of area dependence reported for metals.

\begin{figure}
\includegraphics[width=15cm]{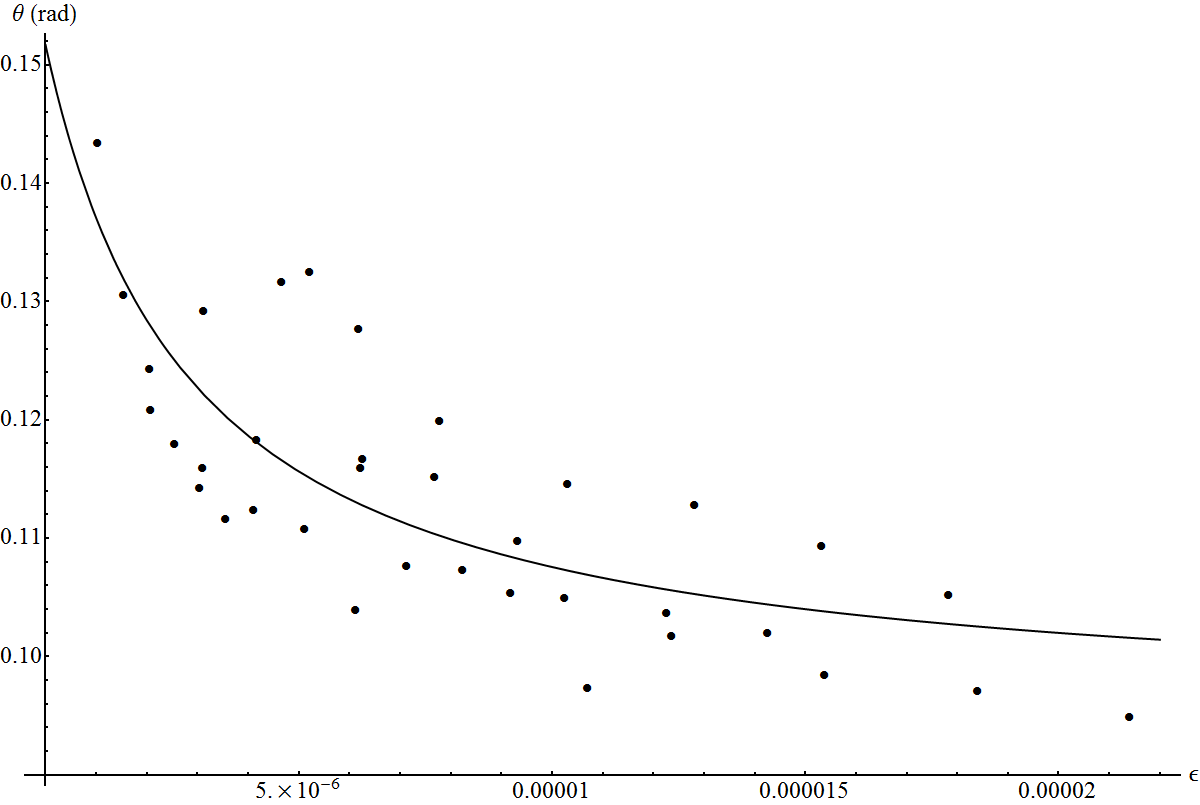}
\caption{Graph of $\theta$ against $\varepsilon$ for various loads and apparent contact areas in the second experiment, the horizontal setup. The curve fitting of Eq. (\ref{fit}) gives $\zeta_0=(7.70\pm1.49)\times10^{-6}$ and $k=2.43\pm0.24$. The adjusted $r^2$ value for the curve fitting is 0.582.}
\label{thetaN3}
\end{figure}

In the second experiment with the horizontal setup, we have a set of six values of $N$ for each of the six different areas, $A$, $A/2$, $A/3$, $A/4$, $A/5$, $A/6$, where $A=0.045$ m$^2$. Similar to the first experiment, $\theta_0=(0.152\pm0.004)$ radians is obtained by extrapolating from the set of points for the largest area. The curve fitting of Eq. (\ref{fit}) for $\theta$ against $\varepsilon$ gives $\zeta_0=(7.70\pm1.49)\times10^{-6}$ and $k=2.43\pm0.24$, with the graph shown in Fig. \ref{thetaN3} (adjusted $r^2$ value of 0.582) \footnote{We do not plot the corresponding graphs of Figs. \ref{thetaN} and \ref{FN} since there are only six points for each area, which would not be very indicative.}. The value of $\zeta_\infty$ can then be calculated from Eq. (\ref{boundary}) to be $1.10\pm0.03$. Supposing once again that $\eta_\infty=0.2$, we then find that $\chi=5.48\pm0.13$ and $\eta_0=(1.40\pm0.27)\times10^{-6}$. Thus, the experiment from the horizontal setup also produces results which agree with the inclined setup, supporting the predictions of our theory. To estimate the uncertainty in this experiment due to sample-to-sample variations across the six different plates, we find for the strain $\varepsilon=6.31\times10^{-6}$, the mean value of the angle is $\theta_{\textrm{mean}}=0.119$ rad with standard deviation of 0.010 rad, indicating a percentage error of $8.3 \%$ or percentage error per sample of $1.4 \%$.

The appearance of the wide scatter in Fig. \ref{thetaN3} is perhaps the repercussion of combining data from plates with (unfortunately) varying degrees of homogeneity, since it is not hard to see the apparent curving characteristic of Eq. (\ref{fit}) especially exhibited by the collection of points below the fitted curve. The softness of teflon also poses a challenge in handling as it is susceptible to mild scratches which destroy the homogeneity \footnote{This is perhaps why chefs are advised to use wooden spatula instead of metal utensils when using a non-stick pan to avoid scraping off the teflon layer.}. The results of the second experiment thus explain why area dependence of static friction is not normally emphasised or highlighted, as such observations could be greatly obfuscated by random errors across multiple plates. These non-Amontons effects would only be strongly manifested in a really precise experiment where state-of-the-art techniques are employed to ensure that different plates have as close a degree of homogeneity as possible throughout the entire measurements process.

A table to summarise these results and compare with our theoretical predictions is provided in Table 1.

\begin{table}
\begin{tabular}{|c|c|c|c|}
  \hline                       
  Quantity				      & Expectation         & \,\,\,1st Expt: Inclined setup\,\,\, & 2nd Expt: Horizontal Setup             \\
  \hline
  $\theta_0$ (rad)		  &	$\sim0.2$ (estimated & $0.178\pm0.001$ 			       			   & $0.152\pm0.004$					     					\\
												& from AFM images)    &																			 &																				\\
  \hline
  $\zeta_0$ 				    &	$>10^{-6}$          &	$(2.40\pm0.18)\times10^{-5}$ 	 	     & $(7.70\pm1.49)\times10^{-6}$ 	  	    \\
  \hline
  $k$ 						      &	$\sim1$             &	0.86$\pm$0.17     					    	   & $2.43\pm0.24$  								       	\\
	\hline
  Adjusted $r^2$ value  &	1								    &	0.923 (See footnote [42].)	 	 	     & 0.582												 	  	    \\
  \hline
  $\zeta_\infty$				&	Eq. (\ref{boundary})&	1.73$\pm$0.22     							  	 & $1.10\pm0.03$									        \\
  \hline
  $\eta_\infty$				  &	Taken to be 0.2\,   & N/A  													       & N/A  														      \\
  \hline
  $\chi$     				    &	$>1$          &	8.64$\pm$1.08  									  	 & $5.48\pm0.13$  						 			      \\
  \hline
  $\eta_0$  				    &	$\sim10^{-6}$		    &	$(2.78\pm0.28)\times10^{-6}$ 			   & $(1.40\pm0.27)\times10^{-6}$      	    \\
  \hline  
  Estimate of $\%$ error& 0 $\%$              &	4.1 $\%$				    			           & 8.3 $\%$    					                  \\
  due to different plates&										&(based on $\varepsilon=1.30\times10^{-6}$)&(based on $\varepsilon=6.31\times10^{-6}$)\\
  \hline  
\end{tabular}
\caption{A summary of the values of the quantities, comparing between the expectations and the two experimental results for teflon-on-teflon. The experimental results reasonably agree with the expectations.}
\label{table1}
\end{table}

\section{Concluding remarks}

In this paper, we have presented a simple, analytic and reversible theory of static friction by modeling the surfaces of contact as being composed of regular elastic smooth microscopic inclines. This theory predicts non-Amontons behaviors, which we have demostrated with teflon (a relatively soft material whose Young's modulus is three orders smaller than metals). The usual Amontons' laws are recovered in the rigid object limit of our theory. We have thus provided a way to summarise and represent the irregular asperities into a useful picture, offering a simplified yet insightful way to think about friction. Our theoretical curve given by Eq. (\ref{fit}) agrees with our experimental data for teflon-on-teflon, with the variance in data accounted by the sample-to-sample variations. These general results and properties of how static friction changes with load and area should still be reasonably reflected in other materials where dry friction effects are dominant. It would hence be interesting to perform future investigations to further examine the applicability of our theory on various other surfaces, as well as extending it to include the description of kinetic friction.

\appendix*
\section{Proof that the differential equation for $\eta$ must be linear}

Consider two blocks of different apparent contact areas $A_A$ and $A_B$ sitting on a large common surface. Suppose the variation of $\eta_A$ and $\eta_B$ for each of these blocks are given by Eq. (\ref{etagen}). A third (combined) block of such material with a total apparent contact area of $A_{AB}=A_A+A_B$ (which can be thought of as putting together those two blocks as a whole) must also be governed by Eq. (\ref{etagen}), since one can perform an experiment with any desired apparent contact area \footnote{In other words, an experimenter using a block with $A_{AB}$, or $A_A$, or even $A_B$ would obtain the same kind of variation for $\eta$ as long as the same material of the same homogeneity is used.}.

The effective compressed area of the combined block is the sum of that of the individual blocks, $A'_{AB}=A'_A+A'_B$. Note also that by definition, $\eta_{A}=A'_A/A_A$, $\eta_B=A'_B/A_B$ and $\eta_{AB}=A'_{AB}/A_{AB}=(\eta_A A_A+\eta_B A_B)/A_{AB}$. Writing out the differential equation for the combined block in terms of the constituent blocks, we have
\begin{eqnarray}
\frac{d\eta_{AB}}{d\varepsilon}=\frac{A_A}{A_{AB}}\frac{d\eta_A}{d\varepsilon}+\frac{A_B}{A_{AB}}\frac{d\eta_B}{d\varepsilon}.
\end{eqnarray}
Since the constituent blocks obey Eq. (\ref{etagen}),
\begin{eqnarray}
\frac{d\eta_{AB}}{d\varepsilon}&=&\frac{A_A}{A_{AB}}k_1(\eta_\infty-\eta_A)+\frac{A_A}{A_{AB}}k_2(\eta_\infty-\eta_A)^2\nonumber\\&\,&+\frac{A_B}{A_{AB}}k_1(\eta_\infty-\eta_B)+\frac{A_B}{A_{AB}}k_2(\eta_\infty-\eta_B)^2+\cdots\\
&=&k_1(\eta_\infty-\eta_{AB})+k_2\left(\eta_\infty^2-2\eta_\infty\eta_{AB}+\frac{\eta_A A'_A+\eta_B A'_B}{A_{AB}}\right)+\cdots\\
&\neq&k_1(\eta_\infty-\eta_{AB})+k_2(\eta_\infty-\eta_{AB})^2+\cdots.
\end{eqnarray}
This shows that whilst the linear term preserves the form of the differential equation for the combined block, the second order term (and similarly for higher order terms) does not. Therefore, the general form of the differential equation for $\eta$ in Eq. (\ref{etagen}) must only contain the linear term:
\begin{eqnarray}
\frac{d\eta}{d\varepsilon}=k(\eta_{\infty}-\eta),
\end{eqnarray}
which is Eq. (\ref{eta}).

\begin{acknowledgments}
We would like to thank Siew Ann Cheong, the first year lab coordinator who endorsed our project and our proposal to setting up the experiments in the Division of Physics and Applied Physics, along with his fastidious effort in reviewing our draft. We wish to also extend our appreciation to Shen Yong Ho and Sunku Sai Swaroop for their support which facilitated the progress of our experiments, as well as Abdul Rahman bin Sulaiman from the mechanical workshop, and also Yuanqing Li for his assistance in producing the AFM images for our polished teflon surfaces.
\end{acknowledgments}

\bibliography{Citation}

\end{document}